\documentclass[AMA,STIX1COL]{WileyNJD-v2}
\usepackage{comment}

\renewcommand{\thefootnote}{\fnsymbol{footnote}}
\articletype{Article Type}%

\received{}
\revised{}
\accepted{}
        
\usepackage{amsfonts}

\newcommand{\sad}{scalar advection-diffusion }
\newcommand{\ourbc}{consistent flux boundary condition }
\newcommand{\ourbcNoSpace}{consistent boundary condition}
\newcommand{\Ourbc}{Consistent flux boundary condition}
\newcommand{\otherbc}{zero diffusive flux boundary condition }
\newcommand{\Peclet}{P\'eclet }
\newcommand{\out}{backflow stabilization }
\newcommand{\Out}{backflow stabilization}
\newcommand{\otherbcNoSpace}{zero diffusive flux boundary condition}

\begin{document}

\title{Numerical Considerations for Advection-Diffusion Problems in Cardiovascular Hemodynamics}


\author[1]{Sabrina R. Lynch}
\author[2]{Nitesh Nama}
\author[3]{Zelu Xu}
\author[4] {Christopher J.Arthurs}
\author[3]{Onkar Sahni}
\author[1,2]{C. Alberto Figueroa}

\authormark{}

\address[1]{\orgdiv{Department of Biomedical Engineering}, \orgname{University of Michigan}, \orgaddress{\state{Michigan}, \country{USA}}}
\address[2]{\orgdiv{Department of Surgery}, \orgname{University of Michigan}, \orgaddress{\state{Michigan}, \country{USA}}}
\address[3]{\orgdiv{Mechanical, Aerospace and Nuclear Engineering}, \orgname{Rensselaer Polytechnic Institute}, \orgaddress{\state{New York}, \country{USA}}}
\address[4]{\orgdiv{School of Biomedical Engineering \& Imaging Sciences}, \orgname{King's College London}, \orgaddress{\state{London}, \country{UK}}}

\corres{\email{srlynch@umich.edu}}

\presentaddress{Computational Vascular Biomechanics Lab, University of Michigan, Ann Arbor, MI USA.}

\abstract[Summary]{Numerical simulations of cardiovascular mass transport pose significant challenges due to the wide range of \Peclet numbers and backflow at Neumann boundaries. In this paper we present and discuss several numerical tools  to address these challenges in the context of a stabilized finite element computational framework. To overcome numerical instabilities when backflow occurs at Neumann boundaries, we propose an approach based on the prescription of the total flux. In addition, we introduce a ``consistent flux" outflow boundary condition and demonstrate its superior performance over the traditional \otherbcNoSpace. Lastly, we discuss discontinuity capturing (DC) stabilization techniques to address the well-known oscillatory behavior of the solution near the concentration front in advection-dominated flows. We present numerical examples in both idealized and patient-specific geometries to demonstrate the efficacy of the proposed procedures. The three contributions discussed in this paper enable to successfully address commonly found challenges when simulating mass transport processes in cardiovascular flows.}



\maketitle

\section{Introduction}

Mass transport of biochemical species plays an important role in numerous cardiovascular pathologies including thrombosis and atherosclerosis. Computational models of mass transport offer the unique capability to study various biochemical processes essential to understand the kinetics of disease progression, but which are otherwise difficult to measure \textit{in vivo}. However, cardiovascular mass transport problems are characterized by highly advective flows (with \Peclet numbers up to $10^7$) that make obtaining an accurate numerical solution challenging. Furthermore, every outlet face of a computational model is an artificial boundary resulting from the arbitrary truncation of a vessel. Therefore, it is necessary to prescribe realistic boundary conditions that result in a stable solution at outlet faces while preserving the accuracy of the solution.

In this work, we present a stabilized finite element framework that incorporates three salient features: (\emph{i}) a \out technique to obtain stable solutions with Neumann outflow boundaries for \sad problems, (\emph{ii}) a \ourbc that minimally disturbs the local physics of the problem on outflow boundaries resulting from the artificial truncation of vessels;
and (\emph{iii}) a front-capturing stabilization technique to regularize the solution with a concentration front in scenarios of high \Peclet numbers.

\textit{Backflow Stabilization}: Neumann conditions have typically been prescribed for outlet boundaries in cardiovascular flows, either through direct imposition of a known traction (i.e. zero or constant pressure condition)\cite{oshima2001finite,perktold1995computer} or, more recently, through the coupling of reduced order models (i.e. lumped parameter networks) of the distal vasculature, which ultimately results in the specification of a time-varying weak traction on the outlet face \cite{Vignon-Clementel2006}. However, Neumann conditions in boundaries exhibiting partial or complete inflow are known to lead to numerical divergence \cite{kim2009augmented,moghadam2011comparison,fouchet2014artificial}. Specifically, prescribing a diffusive flux fails to guarantee stable energy estimates due to the unknown velocity profile at these boundaries \cite{Gravemeier2012}. To mitigate these difficulties associated with flow modeling, several strategies have been proposed including adding a backflow stabilization term to the boundary nodes\cite{Gravemeier2012,bertoglio2018benchmark}, constraining the velocity to be normal to the outlet\cite{moghadam2011comparison}, or using Lagrange multipliers to constrain the velocity profile at all or some of the outlets\cite{kim2009augmented}. A comparison of these strategies determined that \out was the most robust approach with the least impact on both the solution and computational cost\cite{moghadam2011comparison}.


Similar scenarios of numerical instability can arise in \sad systems~\cite{Hughes2005}. Despite the numerous reports on backflow stabilization for flow problems~\cite{kim2009augmented,moghadam2011comparison,bertoglio2014tangential}, these strategies have not been adopted for \sad systems. Instead, to circumvent the numerical instability issues in the presence of backflow, mass transport models have resorted to unphysical approaches such as the imposition of arbitrary Dirichlet boundary conditions at the outlet faces \cite{Hansen2019, Arzani2016}, artificial extensions of the computational domain\cite{Farghadan2019TheTransport} that seek to regularize the flow profile, or an artificial increase in the diffusivity of the scalar~\cite{Biasetti2012, Ford2005VirtualHemodynamics}. In this work, we propose a stabilization method for outlet Neumann boundaries, following the ideas presented by Hughes and Wells~\cite{Hughes2005}. 

\textit{Consistent Flux Boundary Condition}:
While there have been numerous contributions proposing outflow boundary conditions for cardiovascular flow problems~\cite{Vignon-Clementel2006}, little work has been done for the \sad problem. Typically, cardiovascular mass transport models have employed either Dirichlet or Neumann conditions prescribing known scalar concentrations\cite{Hansen2019,Arzani2016} or diffusive fluxes\cite{Leiderman2011,Yazdani2018} at the outlet face, respectively. An alternative choice of boundary condition, henceforth referred to as ``\ourbcNoSpace'' has been shown to provide better error estimates~\cite{griffiths1997no}. This approach relies on calculating the consistent diffusive flux (rather than imposing an arbitrary diffusive flux) that satisfies the weak form of the mass transport equation. To the best of our knowledge, this boundary condition has been thus far unexplored for cardiovascular mass transport problems. In this work, we demonstrate the superior performance of this approach over the traditional zero diffusive flux boundary condition.



\textit{Front-Capturing Stabilization Techniques}: Another important issue concerning simulation of mass transport in cardiovascular applications is the presence of high \Peclet number flows typically found in the large arteries. These advection-dominated flows lead to the development of steep concentration gradients, thereby necessitating the use of stabilization techniques to avoid unphysical oscillations in the numerical solution near the concentration front. To address this issue, several discontinuity capturing methods 
have been proposed~\cite{hughes1986new,codina1993discontinuity,de2001natural}. In this work, we discuss the performance of the discontinuity capturing (DC) stabilization technique, implemented in the context of a streamline upwind Petrov-Galerkin (SUPG) stabilized finite element formulation.


Numerical results are presented in both idealized and patient-specific geometries to demonstrate the efficacy of the proposed numerical procedures. 


\section{Methods}
\subsection{Strong form and boundary conditions}
The strong form of the governing equation for mass transport in a three-dimensional bounded domain $\Omega \subset \mathbb{R}^3$ is given as
\begin{align}
\label{eq: strong_form}
    \frac{\partial c}{\partial t} + \textbf{u} \cdot \nabla c - \nabla \cdot \left(D \nabla c \right) &= r \qquad \text{in} \quad \Omega,
\end{align}
where $c$, $D$, $r$, and $t$ denote the concentration of the scalar, diffusion coefficient, source (or reaction) terms and time, respectively and $\textbf{u}$ is a known, solenoidal velocity field. $\Omega$ is a open set with boundary $\Gamma = \partial \Omega$, such that: 
\begin{align}
\label{eq: boundary_definitions}
    \Gamma = \overline{\Gamma_{D} \cup \Gamma_{N}}, \\
    \Gamma_{D} \cap \Gamma_{N} = \emptyset,
\end{align}
where $\Gamma_D$ and $\Gamma_N$ are the Dirichlet and Neumann partitions of  the boundary $\Gamma$, respectively. 

We consider a further partition of $\Gamma = \overline{\Gamma^\textrm{in} \cup \Gamma^\textrm{out}}$, ~ $\Gamma^\textrm{in} \cap \Gamma^\textrm{out} = \emptyset$ such that: 

\begin{align}
\label{eq: boundary_decomposition}
    \Gamma^\textrm{in}(t) &= \{\textbf{x} \in \Gamma | u_n(\textbf{x},t) \leq 0\}, \\
    \Gamma^\text{out}(t) &= \Gamma - \Gamma^\text{in}(t),
\end{align}
where $u_n$ is the dot product of the velocity with the outward unit normal at the boundary, $\textbf{x}$ is the position vector, $\Gamma^\text{in}(t)$ is the inflow boundary, and $\Gamma^\text{out}(t)$ is the outflow boundary. $\Gamma^\text{in}(t)$ and $\Gamma^\text{out}(t)$  are functions of time owing to the time dependence of the velocity field. In this manuscript, the terms `outlet' and `inlet' are used to refer to spatially fixed positions of boundary faces, while the terms `inflow boundary' and `outflow boundary' are used to refer to regions of the boundary that exhibit inflow and outflow at a given time instant, respectively. Given these definitions, a total of four distinct boundaries can be defined for mass transport problems~\cite{Hughes2005}: 
\begin{align}
\label{eq: Dirichlet_Neumann}
    \Gamma_{\beta}^{\alpha}(t) =\Gamma_{\beta} \cap \Gamma^{\alpha}(t), \quad
    \quad \alpha = \{\text{in},\text{out}\}, \quad 
    \beta = \{\text{D},\text{N}\}
\end{align}
Typically, finite element simulations of mass transport have considered Dirichlet boundary conditions on inlet faces and Neumann boundary conditions on outlet faces. However, this strategy often shows numerical divergence if backflow occurs on the Neumann boundary (e.g., if \ $\Gamma_N^{\text{in}}(t) \neq \emptyset$). Indeed, it has been shown that the prescription of diffusive flux on a $\Gamma_N^{\text{in}} (t)$ boundary fails to guarantee stable energy estimates and therefore leads to numerical divergence~\cite{Gravemeier2012}. Prescribing the total flux on inflow Neumann boundaries $\Gamma_N^{\text{in}}(t)$ mitigates this issue, whereas the diffusive flux can be safely prescribed on outflow Neumann boundaries $\Gamma_N^\text{out}(t)$~\cite{Hughes2005}, viz:

\begin{align}
    D \nabla c \cdot\textbf{n} &= h^\text{out}  \qquad \qquad &&\textrm{on} \qquad\Gamma_N^{\text{out}}(t), \label{eq: outflowNeumann}\\
    -c\textbf{u}\cdot \textbf{n} + D\nabla c \cdot \textbf{n} &= h^\text{in} \qquad \qquad &&\textrm{on} \qquad \Gamma_N^{\text{in}}(t) \label{eq: inflowNeumann}.
\end{align}
Here, $h^\textrm{out}$ and $h^\textrm{in}$ denote the diffusive and total (i.e., advective plus diffusive) flux data, respectively. 

\subsection{Weak form}
The Galerkin weak form for the \sad problem governed by Eq.~\ref{eq: strong_form} is as follows: find $c \in H^1(\Omega)$ such that 
\begin{align}
\label{eq: weak_form}
\int_\Omega \Big[\delta_c \frac{\partial c}{\partial t} +  \delta_c   \textbf{u} \cdot \nabla c + \nabla\delta_c \cdot D \nabla c \Big] \text{d}V- \int_{\Gamma_N} \delta_c \left(D \nabla c \right)\cdot\textbf{n}\text{d}A = \int_\Omega \delta_c r \text{d}V \qquad \qquad \forall \delta_c \in H_0^1(\Omega)
\end{align}
where $\delta_c$ is a weighting function, $H^1(\Omega)$ is a (solution) space of once-differentiable functions satisfying the Dirichlet boundary conditions on $\Gamma_D$, and $H_0^1(\Omega)$ is a (weighting) space of once-differentiable functions vanishing on the Dirichlet boundary $\Gamma_D$.
Since cardiovascular mass transport problems are characterized by high \Peclet number flows, we utilize a SUPG stabilized finite element formulation~\cite{brooks1982streamline}, resulting in the following discrete weak form:
\begin{align}
\label{eq: stabilized_weak_form}
\int_\Omega \Big[\delta_c \frac{\partial c}{\partial t} +  \delta_c   \textbf{u} \cdot \nabla c + \nabla\delta_c \cdot D \nabla c \Big] \text{d}V
&- \int_{\Gamma_N} \delta_c \left(D \nabla c \right)\cdot\textbf{n}\text{d}A \nonumber\\
&+\sum_{i=1}^{n_\textrm{el}} \int_{\Omega_i} \nabla\delta_c \cdot \textbf{u} \tau \mathcal{R} \text{d}V = \int_\Omega \delta_c r \text{d}V \qquad \qquad \forall \delta_c \in H_0^1(\Omega) 
\end{align}
where $n_\textrm{el}$ denotes the total number of elements in the discretized domain, $\Omega_{i}$ is the domain of the $i$-th element, $\tau$ is the stabilization parameter, and $\mathcal{R}$ is the residual given as 
\begin{align}
\label{eq: residual}
\mathcal{R}=\frac{\partial c}{\partial t}+\textbf{u}\cdot \nabla c-D\nabla^2 c - r.
\end{align}
The stabilization parameter $\tau$ is given as
\begin{align}
\tau^{-2} &= \tau_1^{-2} + \tau_2^{-2} + \tau_3^{-2}, \label{eq: tau_def}\\
  \tau_1^{-2} = \left(\frac{\Delta t}{2}\right)^{-2}, \qquad \tau_2^{-2} = \textbf{u}\cdot \textbf{g} \textbf{u}& , \qquad \tau_3^{-2} = 9D^2\textbf{g}:\textbf{g}, \qquad \textbf{g}=\bigg(\frac{\partial \xi}{\partial \textbf{x}}\bigg)^T \frac{\partial \xi}{\partial \textbf{x}}, \label{eq: SUPGParameters}
\end{align}
where $\Delta t$ is the time step size, : is the Frobenius inner product, and $\textbf{g}$ is the metric tensor based on the Jacobian of the mapping between the element coordinates $\mathbf{\xi}$ and the physical coordinates $\textbf{x}$ (e.g. in 1D, $g=4/h^2$). We remark that, without loss of generality, the examples shown in this article all use a zero reaction term and hence no contributions from the reaction terms appear in the above-mentioned stabilization terms.

\subsection{Backflow stabilization and total flux}
\label{sec: backflow}
Using Eqs.~\ref{eq: outflowNeumann} and ~\ref{eq: inflowNeumann}, the Neumann boundary term in Eq.~\ref{eq: stabilized_weak_form} becomes
\begin{align}
    \int_{\Gamma_{N}} \delta_c \left(D \nabla c \right)\cdot\textbf{n} \text{d}A&=\int_{\Gamma_{N}^\text{out}(t)} \delta_c \left(D \nabla c \right)\cdot\textbf{n} \text{d}A+\int_{\Gamma_{N}^\text{in}(t)} \delta_c \left(D \nabla c \right)\cdot\textbf{n} \text{d}A,\nonumber\\
    &=\int_{\Gamma_{N}^\text{out}(t)} \delta_c h^\text{out} \text{d}A+\int_{\Gamma_{N}^\text{in}(t)} \delta_c \left(D \nabla c -c \bf{u}\right)\cdot\textbf{n} \text{d}A+ \int_{\Gamma_{N}^\text{in}(t)} \delta_c c \bf{u}\cdot\textbf{n}\text{d}A, \nonumber \\
    &=\int_{\Gamma_{N}^\text{out}(t)} \delta_c h^\text{out} \text{d}A+\int_{\Gamma_{N}^\text{in}(t)} \delta_c h^\text{in} \text{d}A+ \int_{\Gamma_{N}^\text{in}(t)} \delta_c c \textbf{u}\cdot\textbf{n} \text{d}A. \label{eq: addedterm}
\end{align}
As indicated earlier, diffusive flux Neumann boundary conditions are typically prescribed at outlet faces $\Gamma^\textrm{out}_{N}(t)$. In this scenario, the last term of Eq.\ref{eq: addedterm} (i.e., advective flux) vanishes since $\Gamma_N^{\text{in}}(t)=\emptyset$. However, in cases where $\Gamma_N^{\text{in}}(t) \neq \emptyset$, the total flux must be prescribed and therefore the advective flux term in Eq.~\ref{eq: addedterm} is non-trivial and must be included in Eq.~\ref{eq: stabilized_weak_form} to obtain a stable solution. Previous publications on backflow stabilization for Navier-Stokes problems have introduced a parameter $\beta$ scaling the advective flux contribution\cite{moghadam2011comparison,Gravemeier2012,bertoglio2018benchmark}. However, owing to the lack of mathematical rigor justifying the introduction and choice of such a scaling factor, we do not consider it for the \sad problem.

\subsection{\Ourbc}
\label{sec: ourbc}
The boundary conditions imposed at artificial boundaries generated due to the truncation of a physical domain form a crucial component of the computational model. Since the downstream physics for mass transport applications is often unknown at such artificial boundaries, the task of identifying appropriate conditions that preserve the accuracy of the solution remains challenging.  

While there have been numerous contributions proposing conditions for artificial boundaries in cardiovascular flow problems, little work has been done for the scalar advection-diffusion problem. Typically, cardiovascular mass transport models have employed either Dirichlet or Neumann conditions prescribing scalar concentrations\cite{Hansen2019,Arzani2016} or diffusive fluxes\cite{Leiderman2011,Yazdani2018} at the outlet face as in Eq.~\ref{eq: outflowNeumann}, respectively. Neither of these approaches is ideal, since they assume knowledge of a physical quantity which is typically unknown. While imposing an arbitrary Dirichlet outlet boundary condition renders a stable solution, it severely affects the scalar solution field and has resulted in approaches relying on extending the outflow branches to minimize the impact of such conditions in the region of interest\cite{Hansen2019,Arzani2016}. These approaches also increase the computational cost due to the larger domain. Conversely, a zero diffusive flux condition has been used more sporadically (likely due to the numerical instabilities associated with backflow, as noted above), and, while seemingly less intrusive than a Dirichlet condition, it still fundamentally prescribes an unknown property of the solution field. 

An alternative to this zero diffusive flux condition, proposed in the context of the Navier-Stokes equations by Papanastasiou and Malamataris~\cite{papanastasiou1992new}, is to calculate the diffusive flux that satisfies the weak form of the mass transport equation and iteratively impose it as a boundary condition. This approach amounts to treating the boundary integral in the weak form given by Eq.~\ref{eq: weak_form} as unknown and is particularly useful when analytic or asymptotic techniques cannot predict the physics downstream from the artificial outlets, making it challenging to formulate appropriate boundary conditions at these faces. 
The strategy of leaving an undefined boundary integral could lead to an ill-posed variational form, particularly in diffusion-dominated problems which are more elliptical in nature and thus necessitate specification of conditions on every boundary. Conversely, advection-dominated problems have a stronger hyperbolic behavior and are less likely to be affected by the ill-posedness of this strategy. For a more detailed discussion on the mathematical implications of this approach, we refer the reader to the work of Griffiths~\cite{griffiths1997no} and Renardy~\cite{renardy1997imposing}. 

In the context of the Navier-Stokes equations, these boundary conditions have been referred to using different terminologies such as ``no boundary condition'' or ``free boundary condition''~\cite{griffiths1997no,papanastasiou1992new}. However, here we will refer to them as ``consistent flux boundary condition''.  In this work, we employ both the zero diffusive flux (see Eq.~\ref{eq: outflowNeumann}), and the consistent flux boundary conditions for the \sad problem and compare their performance in preserving the local physics of the computed solution.

\subsection{Discontinuity capturing operator}
In the context of high \Peclet number flows, SUPG stabilized formulations for \sad problems fail to resolve the steep gradients in the solution, resulting in numerical undershoot/overshoot in concentrations near the scalar front. Therefore, in addition to SUPG stabilization, we implemented a discontinuity capturing (DC) operator to resolve steep gradients in the solution~\cite{LeBeau1993SUPGFormulations}. This approach introduces an additional term for each element of the form $\nabla \delta_c \cdot \nu_{DC} \nabla c$ in Eq.~\ref{eq: stabilized_weak_form} similar to the last term on the LHS. $\nu_{DC}$ is defined as

\begin{align}
    \nu_{DC} =\textrm{max}[0,\omega_{DC}]\,\tilde{\textbf{g}},
\end{align}
where $\tilde{\textbf{g}}$ is the contravariant counterpart of the metric tensor introduced in Eq.~\ref{eq: SUPGParameters} and

\begin{align}
    \omega_{DC}=f_{DC}\sqrt{\frac{ \mathcal{R}^2}{\nabla c \cdot \tilde{\textbf{g}} \nabla c }}-\tau \frac{ \mathcal{R}^2}{\nabla c \cdot \tilde{\textbf{g}} \nabla c },
\end{align}
where $f_{DC}=1$ for linear finite elements and $\tau$ is the stabilization parameter defined in Eq.~\ref{eq: tau_def}. 
We remark that the DC scheme makes the weak form of the \sad problem nonlinear. Therefore, the resolution of gradients near the concentration front is obtained at an increased computational expense. For nonlinear scalar problems, however, the increase in computational cost due to the use of DC operator is not high (as discussed later).


\section{Numerical Examples}
In this section, we present numerical results to illustrate the suitability of the proposed computational framework. 
The following applies to all the numerical examples presented in this section: 
\begin{itemize}
    \item A flow solution is first obtained by solving the stabilized Navier-Stokes equations using the cardiovascular hemodynamics modeling environment, CRIMSON (www.crimson.software)\cite{CRIMSONSoftware}. Here, blood is modeled as a Newtonian fluid with a density of $1060~\textrm{kg}/\textrm{m}^3$ and a dynamic viscosity of $0.004~\textrm{Pa}\cdot\textrm{s}$. All walls are modeled as rigid (i.e., homogeneous Dirichlet boundary conditions for the velocity field). 
    \item For the mass transport problems, a constant concentration of $c=10 ~\textrm{mol}/\textrm{mm}^3$ is prescribed at the inlet face and a zero concentration flux boundary condition is applied to all walls. An initial concentration of $c=0 ~\textrm{mol}/\textrm{mm}^3$ is assumed for all mass transport problems.
\end{itemize}

\subsection{Idealized geometries}
To provide a better understanding of specific numerical challenges, we first present results for cases where idealized geometries and problem parameters are chosen to isolate specific numerical challenges. Here, we present results for three specific cases that highlight the effectiveness of the different components of the proposed computational framework. 

\subsubsection{Backflow stabilization}
\begin{figure}[ht!]
\centering
\includegraphics[width=0.6\textwidth]{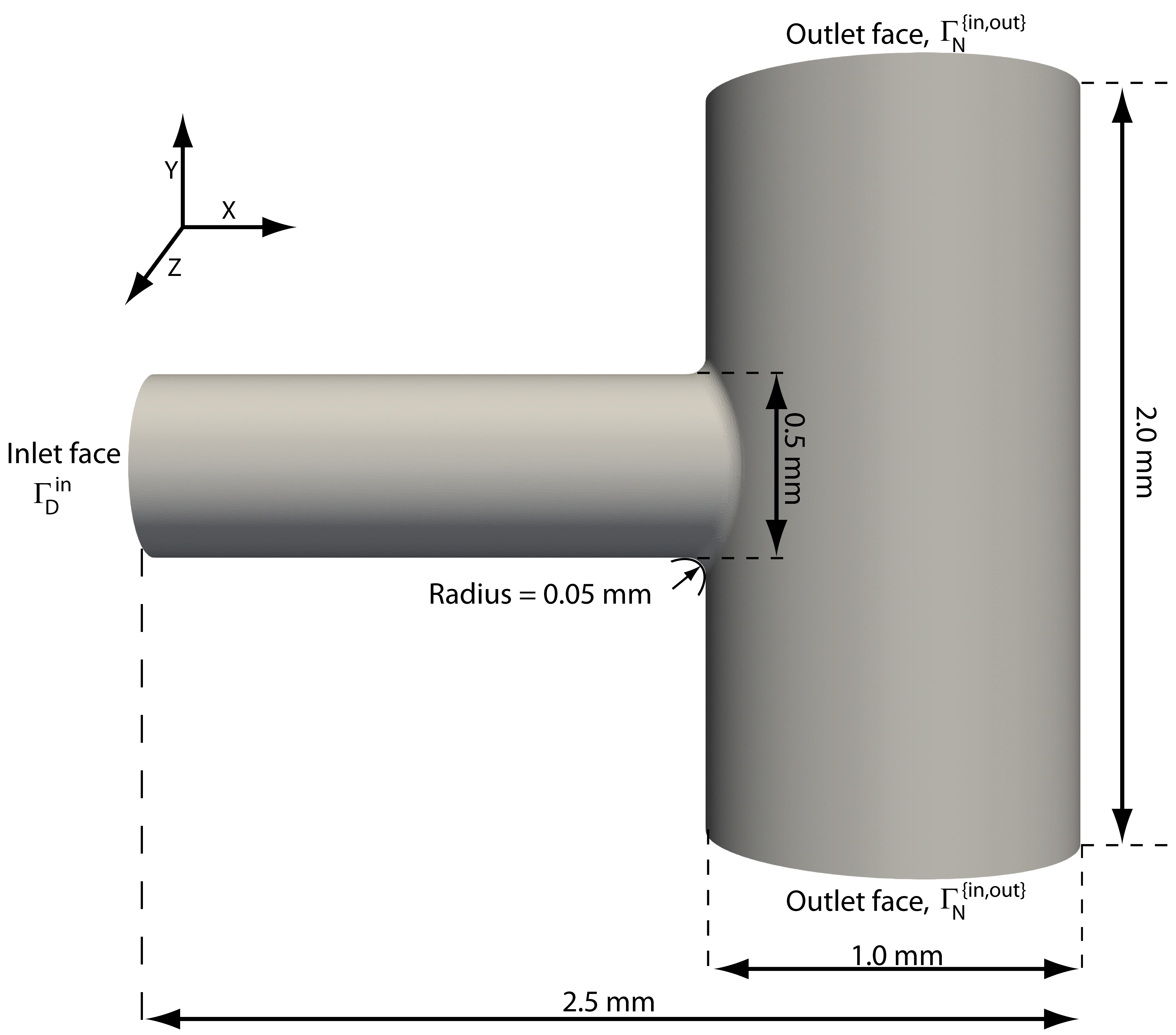}
\caption{3D T-shaped bifurcation model.}
\label{TBifurcation:Geometry}
\end{figure}

To illustrate the numerical issues caused by $\Gamma_N^{\text{in}}(t)$ boundaries, we consider a T-shaped bifurcation  as shown in Figure~\ref{TBifurcation:Geometry}.
The choice of geometry and boundary conditions of this problem leads to partial backflow at the outlet faces even under steady flow conditions. A velocity field was obtained by prescribing a constant inlet flow of $196 ~\textrm{mm}^3/\textrm{s}$, mapped to a parabolic velocity profile, resulting in a maximum velocity of $\textrm{v}_\textrm{max}=2000~\textrm{mm}/\textrm{s}$, a mean velocity of $\textrm{v}_\textrm{mean}=1000~\textrm{mm}/\textrm{s}$, and a Reynolds number $\textrm{Re}_\textrm{mean}=66.25$, based on the mean velocity and the inlet radius. A zero-pressure boundary condition was applied at both outlet faces. 
For the \sad problem, a zero diffusive flux condition was prescribed at the outlet faces. The diffusion coefficient was set to $D = 10^{-2}~\textrm{mm}^2/\textrm{s}$, resulting in a \Peclet number of $\textrm{Pe}_{\textrm{mean}}=2.5\times 10^4$. The domain was discretized using linear tetrahedral elements with characteristic length of $10^{-2}~\textrm{mm}$, resulting in a total mesh size of 11.3 million elements. Simulations were run using a constant time-step size of $\Delta t =10^{-5}~\mathrm{s}$ for $8000$ time steps. 

\begin{figure}[ht!]
\centering
\includegraphics[width=1.0\textwidth]{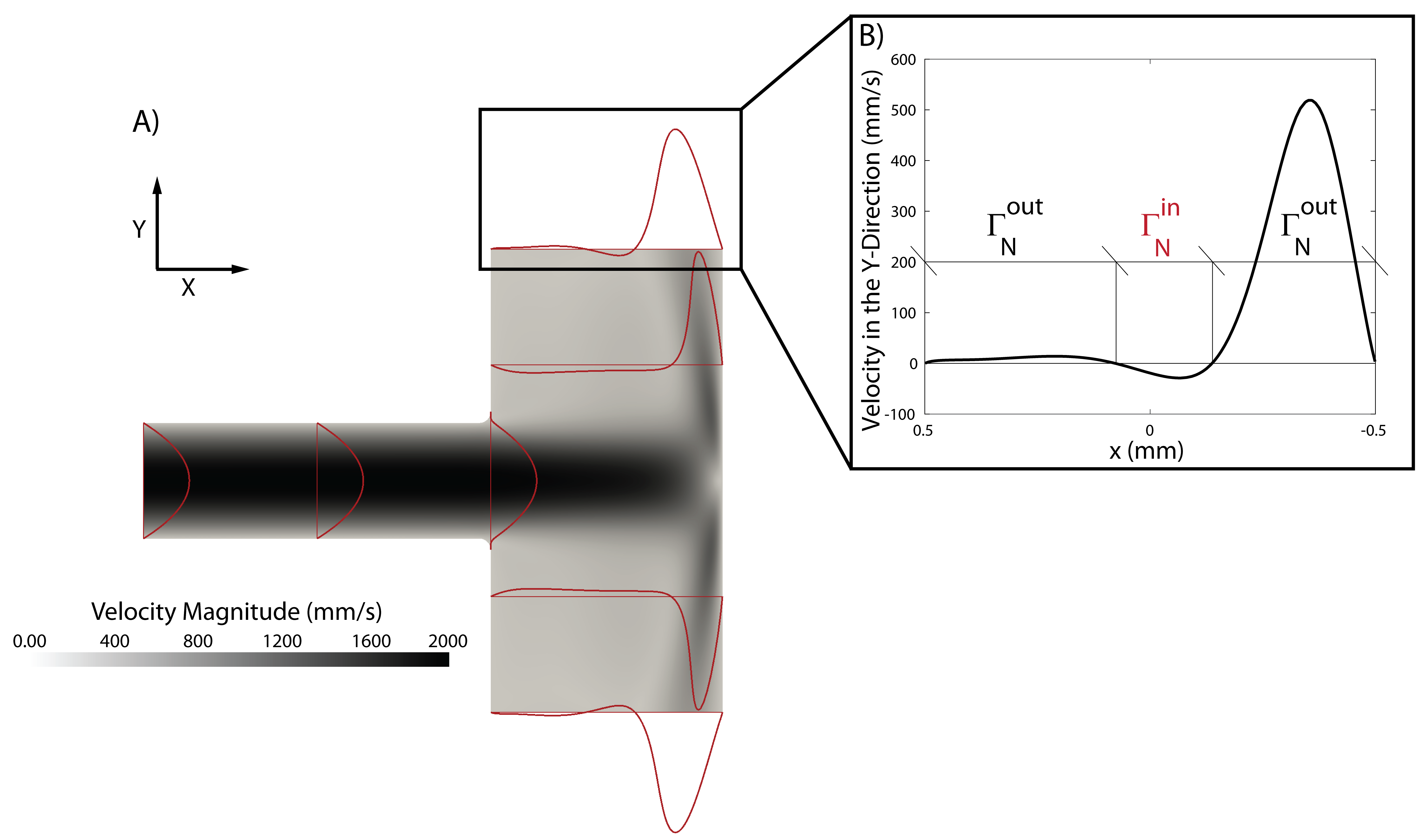}
\caption{A) Velocity contours in the mid-plane of the T-shaped bifurcation. Red lines indicate velocity profiles at discrete number of locations. B) Close-up view of the velocity at the outlets, illustrating backflow in a small segment of the outlet face.}
\label{TBifurcation:Velocity}
\end{figure}
\begin{figure}[ht!]
\centering
\includegraphics[width=1.0\textwidth]{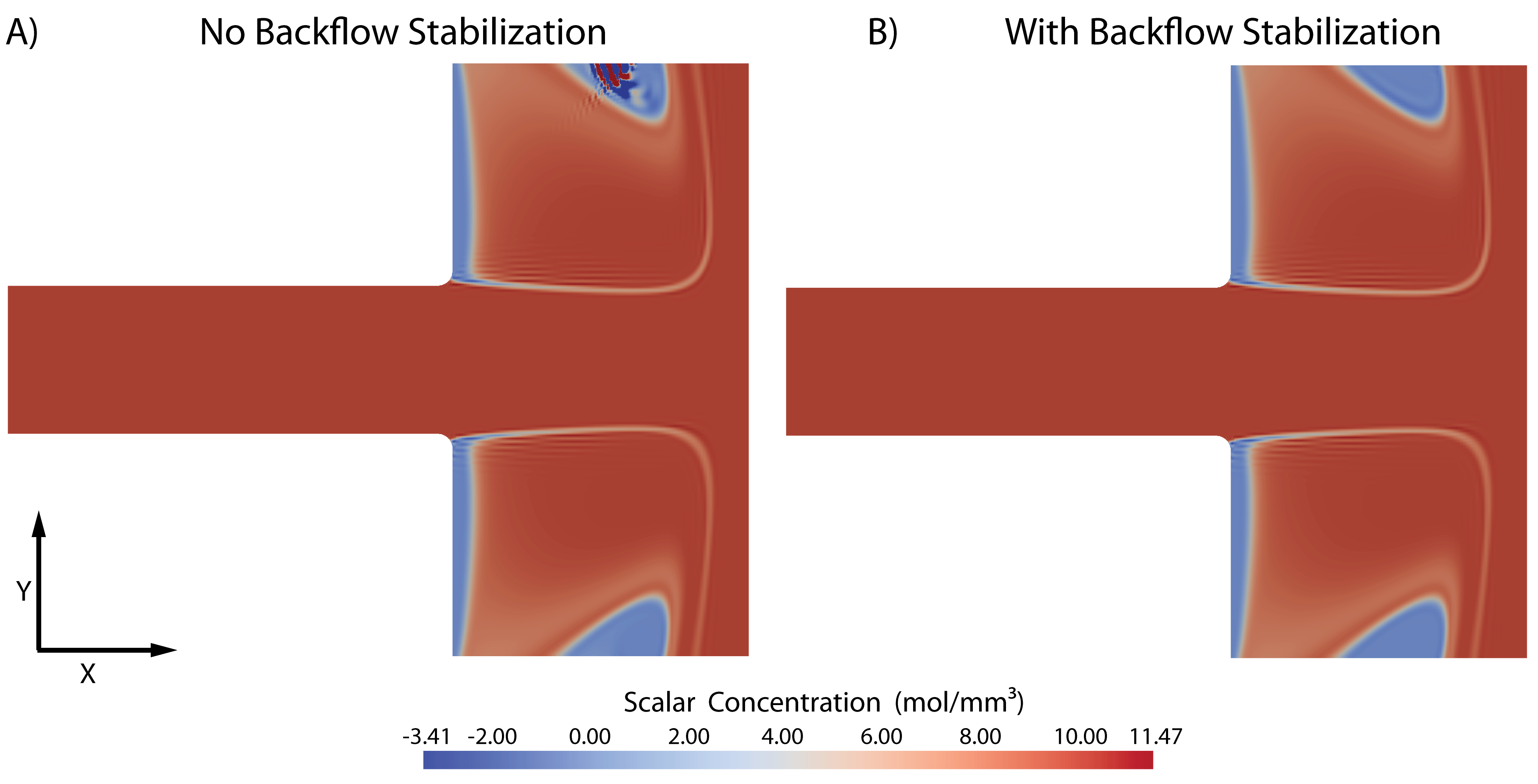}
\caption{Scalar contours at time = 0.036s in the T-Bifurcation. A) No \out resulting in an unstable solution, and B) With \out resulting in a stable solution in the presence of backflow.}
\label{TBifurcation:Divergence}
\end{figure}

Figure~\ref{TBifurcation:Velocity}(A) shows the velocity field plotted at the mid-plane of the T-shaped bifurcation perpendicular to the $Z$ direction. The velocity profiles at the outlet faces exhibit backflow, Figure~\ref{TBifurcation:Velocity}(B). Figure~\ref{TBifurcation:Divergence}(A) shows the solution for the scalar concentration at $t=0.036~\textrm{s}$, obtained without backflow stabilization. The solution presents strong numerical artifacts on the top outlet boundary, corresponding to the $\Gamma_N^{\text{in}}(t)$ region of the outlet boundary. These numerical artifacts eventually lead to divergence of the simulation and also appear in the bottom outlet boundary over time. In contrast, the proposed backflow stabilization technique yields a stable solution as shown in Figure~\ref{TBifurcation:Divergence}(B). We would like to point out that the two numerical solutions in Figure~\ref{TBifurcation:Divergence}(A) and Figure~\ref{TBifurcation:Divergence}(B) exhibit spurious oscillations in the interior of the domain, these will be addressed in Section~\ref{sec:DCIdealized}.


\subsubsection{\Ourbc}
\label{sec: ourbcIdealized}

In this example, we compare the behavior of the consistent flux and the zero diffusive flux boundary conditions. We consider two cylindrical domains of diameter $d=1.0~\textrm{mm}$
and lengths $l_1=10~\mathrm{mm}$ and $l_2=5~\mathrm{mm}$, respectively. For each cylindrical domain, we apply both types of boundary conditions, rendering a total of four different scenarios. The ultimate goal of this test is to examine the impact of the boundary conditions on the scalar field, specifically by comparing solutions at the outlet of the shorter cylinder with solutions at the mid-section of the longer cylinder, which are taken as the ``reference solution". A desirable feature of the boundary condition is to minimize the impact on the scalar field, given that these conditions are typically prescribed on ``artificial" boundaries (i.e., arbitrary truncations of a branch). 

Steady flow field solutions were obtained by prescribing a constant flow rate of $196~\textrm{mm}^3/\mathrm{s}$, mapped to a parabolic velocity profile resulting in a maximum velocity of $\textrm{v}_\textrm{max}=500~\textrm{mm}/\textrm{s}$, a mean velocity of $\textrm{v}_\textrm{mean}=250~\textrm{mm}/\textrm{s}$ and a Reynolds number of $\textrm{Re}_\textrm{mean}=33.125$. A zero-pressure boundary condition was applied at the outlet face. 
For the mass transport problem, a constant value of diffusivity was adopted, $D = 10^{2}~\textrm{mm}^2/\textrm{s}$, resulting in \Peclet number of $\textrm{Pe}_{\textrm{mean}}=1.25$. This relatively low \Peclet number was chosen because it amplifies the differences between zero diffusive and consistent flux boundary conditions. Simulations were run using a constant time-step size of $\Delta t =10^{-4}~\mathrm{s}$ for $8000$ time steps.

\begin{figure}[h!]
\centering
\includegraphics[width=1.0\textwidth]{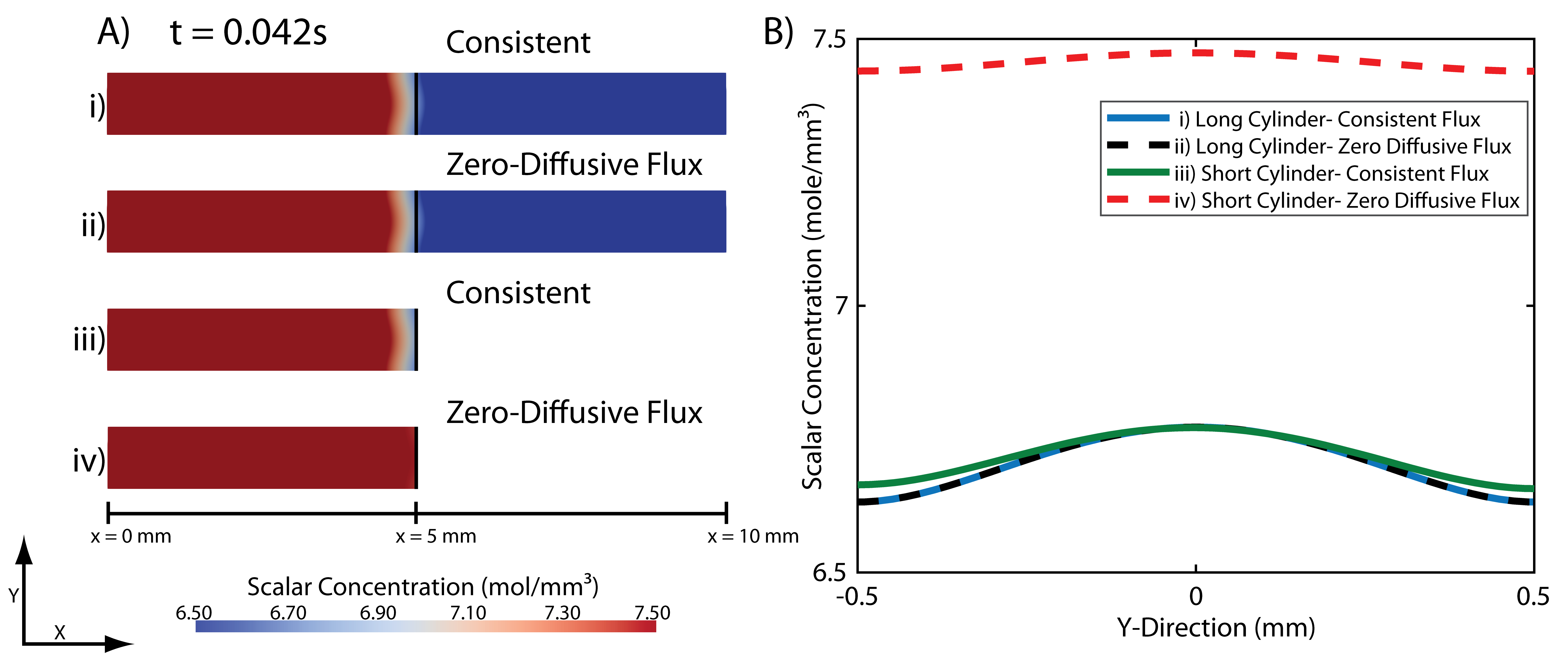}
\caption{A) Scalar contours for four different cylinders. From top to bottom: (i) $10~\textrm{mm}$ cylinder with the consistent flux outflow boundary condition, (ii) $10~\textrm{mm}$ cylinder with a zero Neumann outflow boundary condition, (iii) 5mm cylinder with the consistent flux  boundary condition, and (iv) $5~\textrm{mm}$ cylinder with a zero Neumann boundary condition. B) Line plot showing scalar concentration across the cylinder at $\mathrm{X}=5~\textrm{mm}$ for cases i-iv.}
\label{Cylinder:BCs}
\end{figure}

Figure~\ref{Cylinder:BCs}(A) shows scalar concentration contours at $t=0.042$ s. for the long (i-ii) and short (iii-iv) cylinders, respectively. Figure~\ref{Cylinder:BCs}(B) shows scalar concentration profiles at the mid-section of the long cylinder (i-ii) and the outlet face of the short cylinder (iii-iv). For the long cylinder, the solutions overlap each other, indicating that at this location far away from the boundary, and this point in time, there is no noticeable difference between solutions obtained with either boundary condition. For the short cylinder, however, there is a noticeable difference between the solutions obtained with the zero diffusive and consistent flux boundary conditions. Taking the solution in the long cylinder as the ``reference solution'', it can be observed that the zero diffusive flux condition renders $10\%$ larger scalar concentration values. In contrast, the consistent flux boundary condition yields scalar values much closer to the true solution, overestimated by just $0.015\%$. These results illustrate the superior performance of the \ourbc in preserving the local physics of the numerical solution near artificial boundaries.

\subsubsection{Discontinuity capturing operator}
\label{sec:DCIdealized}
Having addressed numerical issues concerning outlet boundaries, we now focus our attention to spurious oscillations in the numerical solution around the concentration front within the computational domain. We consider the flow solution for the shorter cylindrical domain ($l_2=5~\mathrm{mm}$) described in the previous section. A smaller diffusion coefficient $D = 10^{-2}~\textrm{mm}^2/\textrm{s}$ was adopted, resulting in a higher \Peclet number of $\textrm{Pe}=1.25\times10^4$, that is of practical interest and exhibits spurious oscillations. A \ourbc was prescribed at the outlet face. Same mesh and time step size were used as in the previous section.
 
\begin{figure}[h!]
\centering
\includegraphics[width=0.8\textwidth]{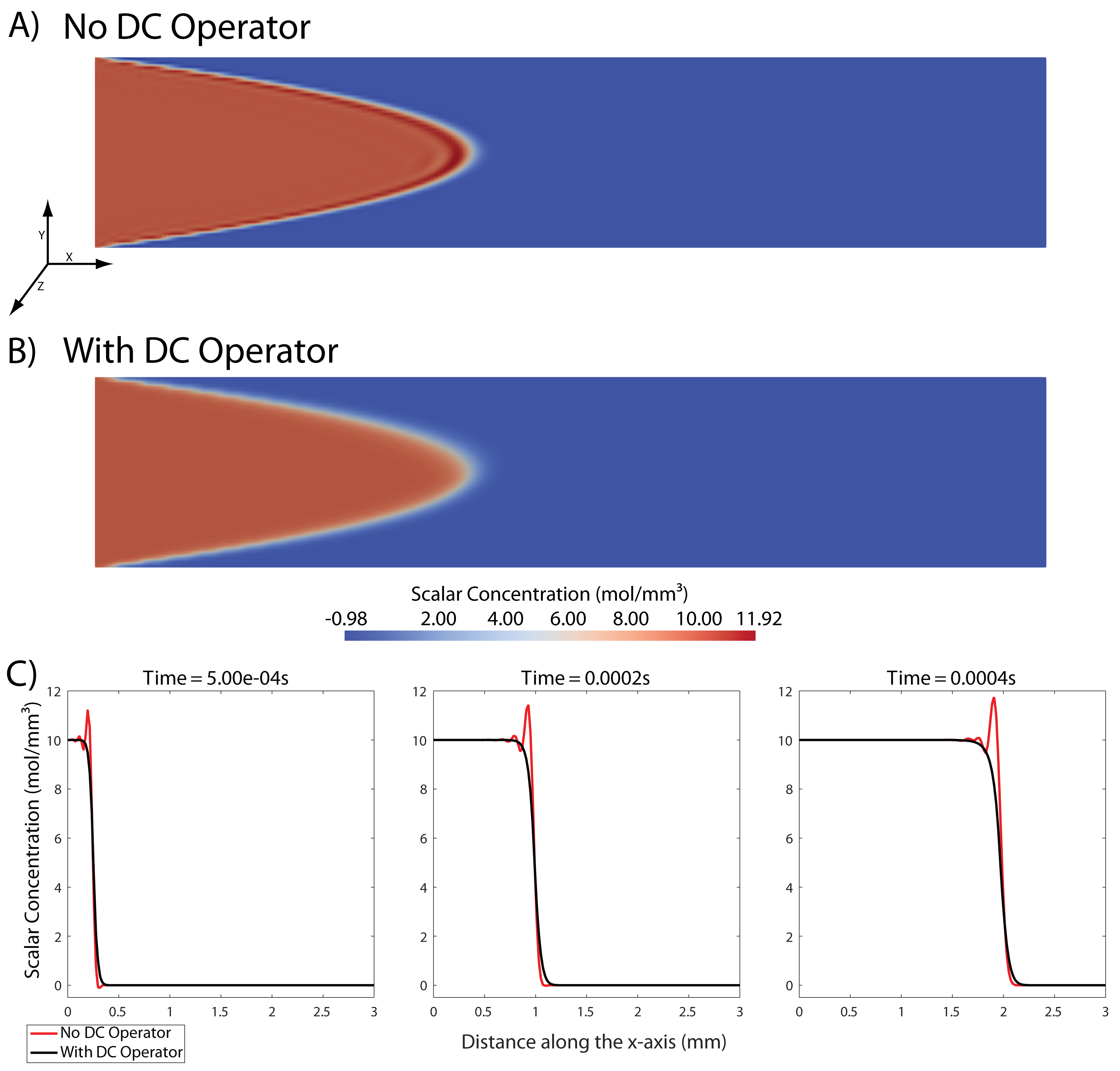}
\caption{A) Scalar contours of $5\mathrm{mm}$ cylinder without the DC Operator at $t=4\times10^{-3}~\textrm{s}$. B) Scalar contours with the DC Operator at $t=4\times10^{-3}~\textrm{s}$. C) Scalar concentration along the center of the cylinder with and without DC Operator at three instances in time: $t=5\times10^{-4}~\textrm{s}$, $t=2\times10^{-3}~\textrm{s}$ and $t=4\times10^{-3}~\textrm{s}$.}
\label{Cylinder:DCMallet}
\end{figure}

Figures~\ref{Cylinder:DCMallet}(A) and Figure~\ref{Cylinder:DCMallet}(B) show concentration contours obtained without and with the DC operator, respectively. Undershoot/overshoot in the numerical solution is apparent near the wavefront of the scalar field when no DC operator is used. These oscillations result in unphysical negative scalar concentrations ($-0.98 \,\mathrm{mol}/\mathrm{mm}^3$) as well as in values higher than those imposed at the inlet ($11.92\,\mathrm{mol}/\mathrm{mm}^3$). Figure~\ref{Cylinder:DCMallet}(C) shows plots of the scalar concentration along the centerline of the cylinder at different times. It can be observed that spurious oscillations begin in the numerical solution without the DC operator (red line) within the first five time steps ($\mathrm{t} = 0.0005~\mathrm{s}$) and increase in magnitude with time. In contrast, the use of the DC operator (black line) results in smooth solution profiles for all times. 


\subsection{Patient-specific geometry}
\label{sec:PtSpecificGeom}
\begin{figure}[h!]
\centering
\includegraphics[width=0.8\textwidth]{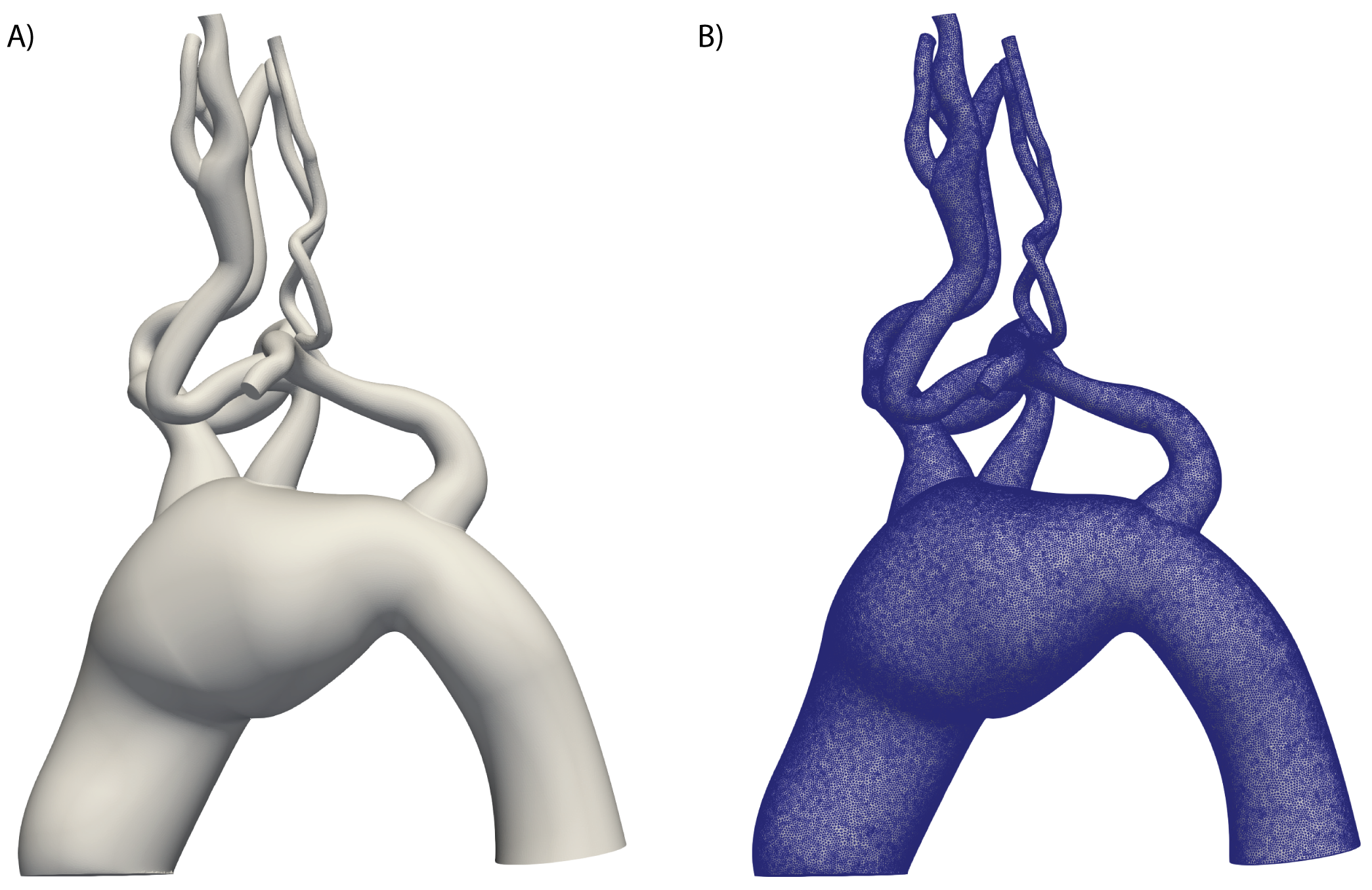}
\caption{A) 3D geometric model reconstructed from CTA image data. B) Computational mesh used in all patient-specific simulations. Both the geometric model and computational mesh were created using CRIMSON. }
\label{AA:Geometry}
\end{figure}

Having demonstrated the capabilities of the stabilized computational framework in idealized geometries under steady flows, we now shift our focus to a patient-specific geometry of a human thoracic aortic aneurysm under periodic flow conditions\cite{VanBakel2018}. The aortic geometry was built from computed tomography angiography (CTA) image data using the cardiovascular hemodynamic modeling environment CRIMSON\cite{CRIMSONSoftware}. Figure~\ref{AA:Geometry} shows the computational domain comprised of the ascending aorta and $9$ outlet branches. The aortic geometry was discretized into $6.2$ million linear tetrahedral elements and $1.1$ million nodes. An echocardiography-derived periodic flow waveform (with time period $T=0.91\,\mathrm{s}$) mapped to a parabolic velocity profile was imposed at the aortic inflow, resulting in a maximum Reynolds number of approximately $\mathrm{Re_{max}}=2.1 x 10^3$.
Three-element Windkessel models \cite{Vignon-Clementel2006} were prescribed at each outlet face, representing the behavior of the distal vascular beds (numerical values given in Supplementary Material). Cycle-to-cycle periodicity was achieved after running the flow problem under rigid wall assumptions for four cardiac cycles, corresponding to a physical time of $t=3.64\,\mathrm{s}$. Subsequently, the \sad equation was solved, assuming a zero concentration initial condition, a constant Dirichlet inlet boundary condition of $c=10\,~\mathrm{mol}/\mathrm{mm}^3$, and zero total flux boundary conditions at the vessel walls for $t>3.64\,\mathrm{s}$. Simulations were run using a constant time step size of $\Delta t = 10^{-4}\,\mathrm{s}$. 

\subsubsection{Backflow stabilization}
\begin{figure}[ht!]
\centering
\includegraphics[width=1.0\textwidth]{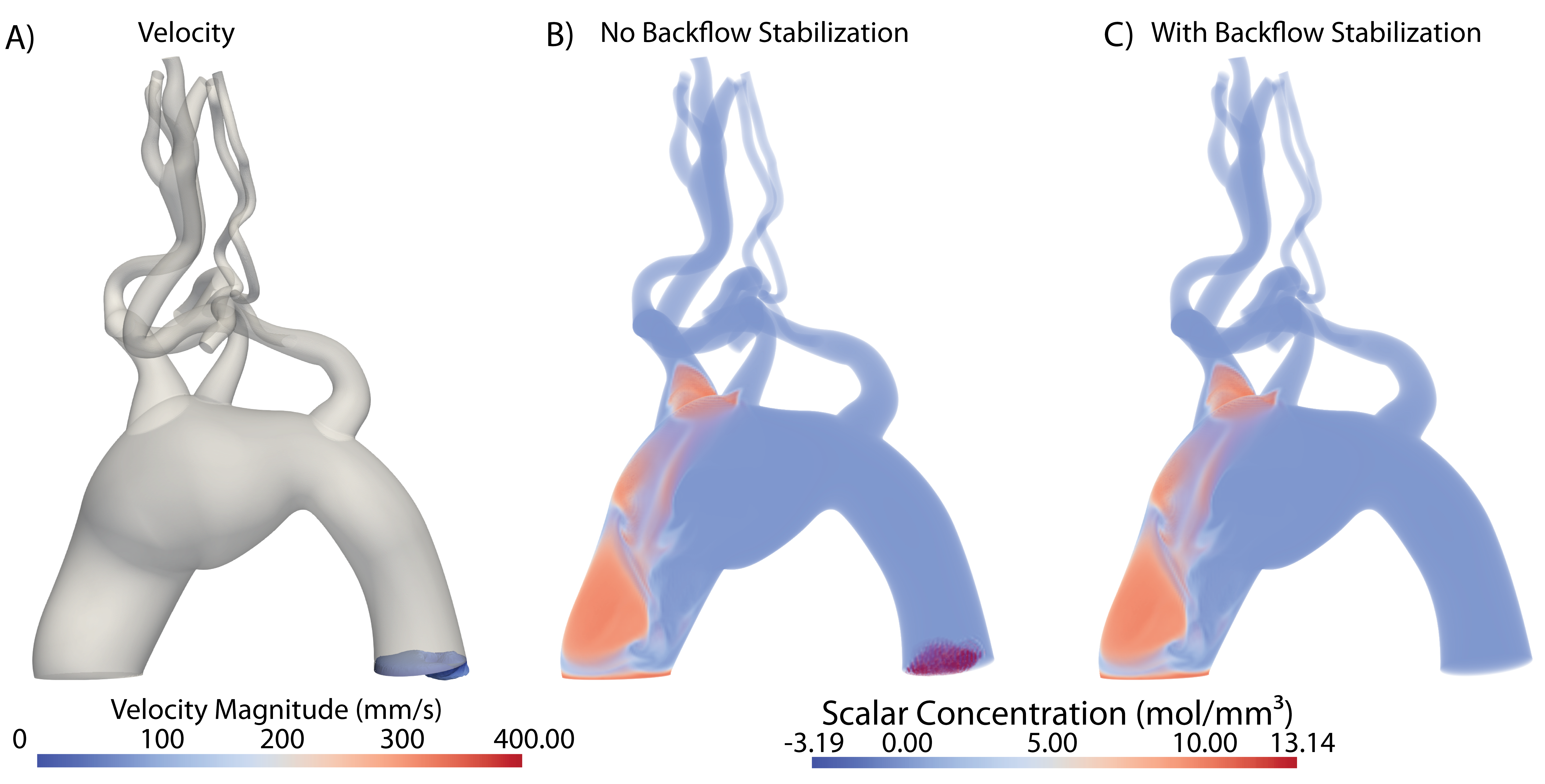}
\caption{A) Computational domain with thoracic aortic aneurysm showing flow reversal at the descending thoracic aorta outlet. Surface contours of scalar concentration at time $t = 4.39\,\mathrm{s}$. B) Without scalar \out numerical instability is observed at the thoracic aorta outlet that proceeds to pollute the scalar domain. C) With \out a stable scalar solution is obtained in the presence of backflow at an outlet.}
\label{AA:Diverging}
\end{figure}

We first studied the issue of numerical instabilities on inflow Neumann boundaries ($\Gamma_N^{\text{in}}$). A constant diffusion coefficient, $D = 1.0\,\mathrm{mm}^2/\mathrm{s}$ was used, resulting in a maximum \Peclet number of $\textrm{Pe}_\text{max}=8.0 x 10^3$ at the inlet face. Zero diffusive flux boundary conditions were prescribed on each outlet face. Figure \ref{AA:Diverging}(A) shows a 3D warp of the velocity profile at the aortic outlet boundary in mid-diastole ($t=4.39\,\mathrm{s}$). Flow reversal is apparent on this boundary at this point in time. Using the standard zero diffusive flux boundary condition without \out leads to instabilities in the numerical solution and eventual divergence, see Figure~\ref{AA:Diverging}(B). Figure~\ref{AA:Diverging}(C) shows the corresponding stable scalar concentration solution obtained with the inclusion of \Out. There was no significant difference in computational cost between solutions obtained with and without backflow stabilization.

\subsubsection{\Ourbc}
\begin{figure}[h!]
\begin{center}
\includegraphics[width=1.0\textwidth]{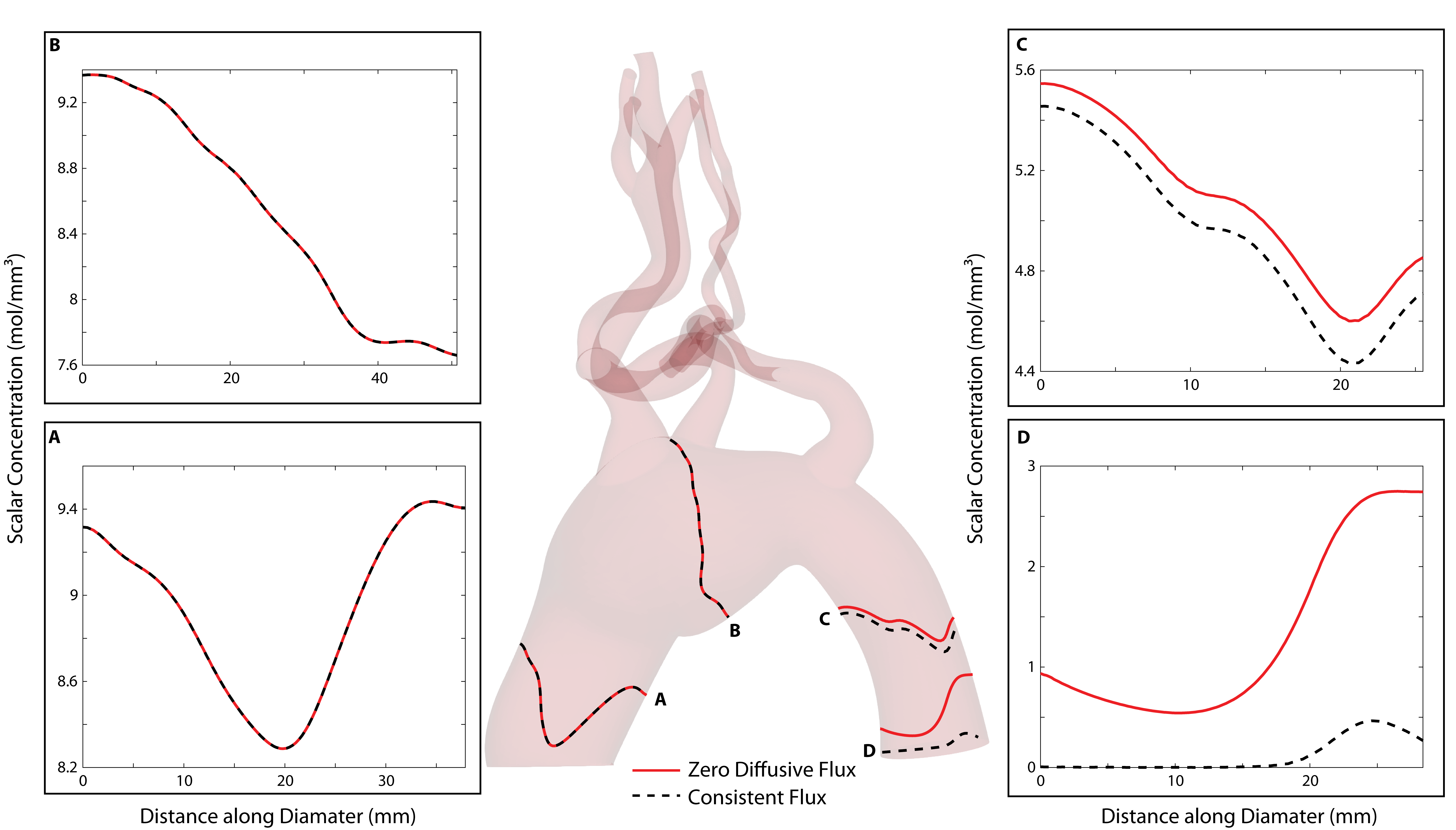}
\end{center}
\caption{Patient-specific simulations in a human thoracic aneurysm were run with both a \otherbc (solid red line) and a \ourbc (dashed black line). Comparisons of the scalar profile across the diameter of the model is shown at four locations (A-D). Results show that close to the inflow the scalar profile across the aorta is the same for both boundary conditions at outlet faces. After the thoracic aneurysm the scalar profiles begin to differ and the greatest differences are observed near the primary outlet face.}
\label{AA:BCs}
\end{figure}

We next studied the performance of the consistent flux versus the zero diffusive flux boundary condition. In both cases, a constant value of diffusion coefficient $D = 10^2\,\mathrm{mm}^2/\mathrm{s}$ was used resulting in a maximum \Peclet number of $\textrm{Pe}_\textrm{max} = 80$ at the inlet face. Figure~\ref{AA:BCs} shows the geometric model with four arbitrary locations A-D along the aorta. Scalar concentration profiles obtained with both boundary conditions at $t=6.55\,\mathrm{s}$ for locations A-D are given. The concentration profiles show substantial variations along the cross section of the vessel for each location, highlighting the contribution of the advection to the concentration profile. For instance, panels A and B show larger values of concentration along the outer curvature of the aorta, where the velocity field is larger.  Panels A and B also show close agreement between the solutions obtained with each boundary condition. In contrast, panels C and D show clear differences between the scalar concentration profiles, with discrepancies between solutions increasing in locations closer to the outlet boundary. Location D shows substantial differences in numerical values and concentration profiles between the two solutions. These results highlight the intrusiveness of the zero diffusive flux boundary condition, particularly in regions of the computational domain near the outlet boundaries. There was no significant difference in computational cost between solutions obtained with the zero diffusive or consistent flux boundary conditions.

\subsubsection{Discontinuity capturing operator}

\begin{figure}[h!]
\centering
\includegraphics[width=1.0\textwidth]{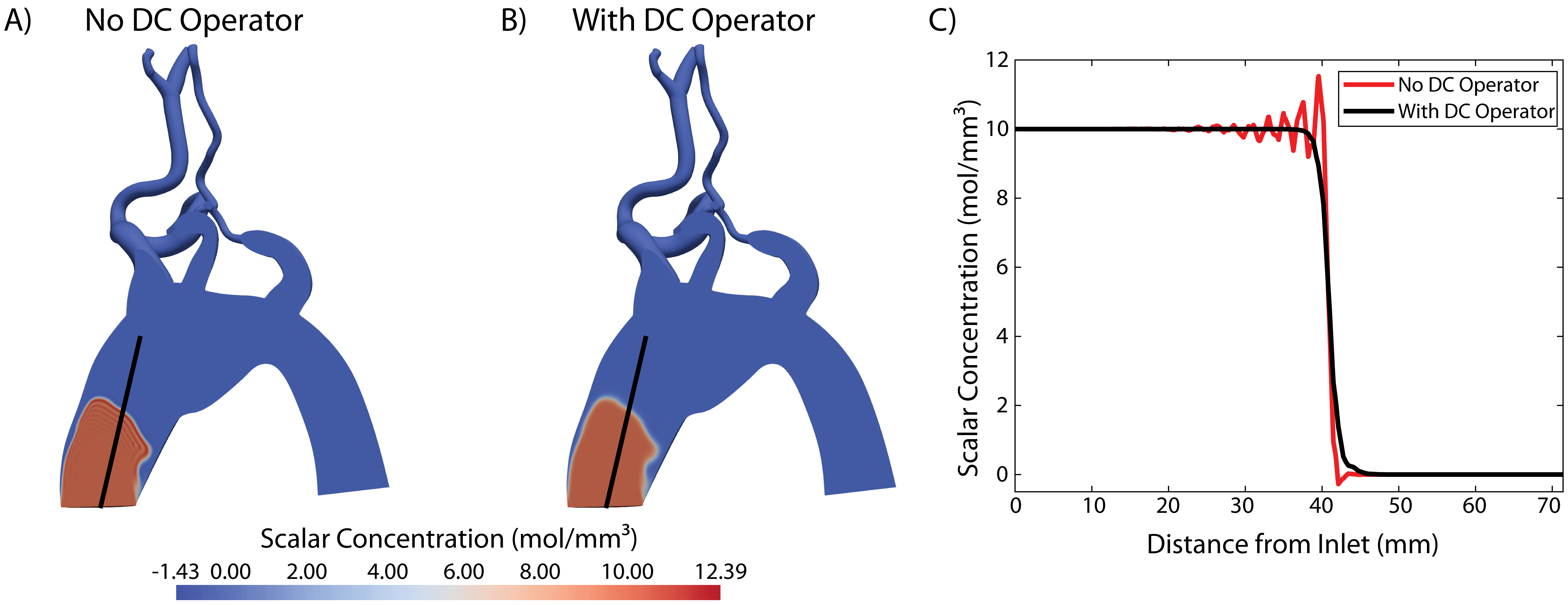}
\caption{Scalar concentration contours at $t=4.04\,\mathrm{s}$ obtained without (A) and with the DC operator (B), respectively. Oscillations in the scalar solution can be seen near the wavefront in (A); a smooth concentration solution can be seen in (B). The lines (in A and B) indicate the location where the scalar concentration profiles are shown in (C). The use of the DC operator effectively avoids the overshoot/undershoot phenomena seen in the simulation with no DC.}
\label{AA:DC}
\end{figure}

In this last example, we demonstrate the efficacy of the DC operator in resolving spurious oscillations in the scalar concentration solution. A constant diffusion coefficient, $D = 1.0\,\mathrm{mm}^2/\mathrm{s}$ was used, resulting in a maximum \Peclet number of $\mathrm{Pe}_\textrm{max} = 8.0 x 10^3$ at the inlet face. Consistent flux boundary conditions were prescribed on each outlet face. Figure~\ref{AA:DC}(A) and (B) show the concentration contours at $t=4.04\,\mathrm{s}$ obtained without and with the DC operator, respectively. Numerical undershoot/overshoot is observed near the wavefront of the scalar field when no DC operator is used.  Figure~\ref{AA:DC}(C) shows a comparison between the two scalar concentration solutions, plotted along an arbitrary line passing through the concentration wavefront. It can be observed that the solution without the inclusion of the DC operator is characterized by spurious oscillations near the concentration wavefront. These oscillations result in unphysical (negative) minimum ($-1.43\,\mathrm{mol}/\mathrm{mm}^3$) and maximum ($12.39\,\mathrm{mol}/\mathrm{mm}^3$) values of concentration. In contrast, the solution obtained with the inclusion of the DC operator shows always positive, smoothly varying scalar concentrations across the wavefront, devoid of any spurious oscillations. 

\section{Discussion and Conclusions}
Transport problems are of paramount importance in studying cardiovascular pathologies. Diseases such as intimal hyperplasia, atherosclerosis and thrombosis are all directly affected by complex transient hemodynamics as well as the transport of numerous chemical species and proteins \cite{Hathcock2006FlowThrombosis, Tarbell2003MassAtherosclerosis, Coppola2009ArterialSignificance, Kaazempur-Mofrad2005MassModels, Perktold2002FluidAnastomoses}. Modeling mass transport in cardiovascular systems presents numerical challenges due to the inherently complex and time-dependent flow patterns and vessel geometries. This complexity is further compounded by the large range of \Peclet numbers found in cardiovascular flows. The primary aim of this work is to present a stabilized computational framework to study 3D, transient cardiovascular mass transport problems. This includes the identification of appropriate boundary conditions that allow for physiologic flow reversal as well as use of stabilization techniques to avoid spurious numerical oscillations in the computed concentration field. 

A number of previous computational studies have employed modeling assumptions or simplifications that are difficult to justify in cardiovascular mass transport problems. These have included unphysiologically large values of diffusion coefficients to lower the \Peclet number \cite{Biasetti2012, Ford2005VirtualHemodynamics}; unrealistic extension of model branches to regularize velocity profiles near outlet faces\cite{Hansen2019, Arzani2016}; and the prescription of arbitrary concentration or flux values at outlet boundaries\cite{Hansen2019, Arzani2016} with a significant influence on the computed solution in regions of interest. These simplifications, always resulting from limitations in the numerical approach, severely limit the applicability of such models for general cardiovascular mass transport studies. 

The computational framework for cardiovascular mass transport presented in this work has three salient features. Firstly, we have presented a \out  strategy to obtain stable numerical solutions in the presence of flow reversal at outlet boundaries. Secondly, we have introduced a `consistent flux boundary condition' and have demonstrated its superiority over the typically used \otherbc in preserving the local physics of the numerical solution, particularly in cases of low \Peclet numbers. Lastly, this framework employs SUPG and DC formulations to resolve steep concentration gradients in mass transport characterized by high \Peclet numbers. 

We have demonstrated the application of this framework in two different sets of geometries. Firstly, we chose idealized geometries with steady flow conditions to allow for a clear interpretation of different numerical challenges and to illustrate the efficacy of the various stabilization techniques reported in this work. The second set of application examples considered a patient-specific model of a human aortic aneurysm under pulsatile flow conditions. This example illustrates the applicability of the framework to complex cardiovascular mass transport problems.

Figure~\ref{TBifurcation:Divergence}(A) demonstrates the issue of numerical divergence in simulations with backflow at Neumann boundaries. The problem was set up in such a way that even under steady flow conditions, flow reversal occurred on some fraction of the outlet boundaries ($\Gamma_N^{\text{in}}$). For a diffusion coefficient $D = 10^{-2}~\textrm{mm}^2/\textrm{s}$ and \Peclet number of $\textrm{Pe}_{\textrm{mean}}=2.5\times 10^4$, simulations diverged if no backflow stabilization was utilized. Following concepts used for stabilization of outlet boundaries in flow problems, the backflow stabilization condition adds an advective component to the diffusive boundary flux. When running experiments with lower \Peclet numbers (i.e., with larger diffusion coefficients), stable solutions can be obtained even without any backflow stabilization. This is expected since the contribution of the advective flux to the total flux decreases with smaller \Peclet numbers. This observation explains that previous studies could report stable solutions without using stabilized outlet boundary conditions for mass transport problems~\cite{Biasetti2012, Ford2005VirtualHemodynamics}. However, an artificial increase in diffusion coefficient changes the physics of the problem entirely.

In Section~\ref{sec: ourbcIdealized}, we studied the performance of the ``consistent flux'' versus standard zero diffusive flux boundary conditions in a short cylindrical geometry, and compared the results against reference solutions obtained in an extended cylindrical geometry. Simulations demonstrated the superiority of the \ourbc in preserving the local accuracy of the solution near the outlet boundary (Figure~\ref{Cylinder:BCs}), albeit a marginal difference relative to the reference truth solution are still noticeable. Important to note, the diffusion coefficient used in this example was increased to $D = 10^{2}~\textrm{mm}^2/\textrm{s}$, leading to a smaller \Peclet number of  $\textrm{Pe}_{\textrm{mean}}=10$. Simulations run with larger \Peclet numbers show smaller differences between consistent and zero flux boundary condition results.

In Section~\ref{sec:DCIdealized}, the performance of the DC operator to stabilize oscillations in the wavefront of the scalar field for high \Peclet numbers was studied. A diffusion coefficient $D = 10^{-2}~\textrm{mm}^2/\textrm{s}$, rendering a \Peclet number $\textrm{Pe}=10^4$ was considered. Figure~\ref{Cylinder:DCMallet} illustrates that without the DC operator, overshoot/undershoot in the numerical solution occurs at the wavefront, resulting in  unphysical negative scalar concentrations ($-0.98\,\mathrm{mol}/\mathrm{mm}^3$) as well as in concentration values higher than those imposed at the inlet ($11.92 \,\mathrm{mol}/\mathrm{mm}^3$). The DC operator eliminates the spurious oscillations, rendering a smooth solution without unphysical negative concentrations. 

The performance of the three formulations was then tested in a patient-specific aortic aneurysm geometry under pulsatile conditions, see Section~\ref{sec:PtSpecificGeom}. The \out produced stable results in the presence of significant flow reversal (Figure~\ref{AA:Diverging}). The \ourbc showed substantial differences in scalar concentration profiles compared to the \otherbcNoSpace, specfically in regions near the outlet boundary, see Figure~\ref{AA:BCs}. Lastly, the DC operator rendered smooth concentration profiles near the wavefront of the solution for high \Peclet number transport, see Figure~\ref{AA:DC}.   

While the different formulations presented in this work provide a set of robust tools to enable simulation of cardiovascular mass transport under realistic geometries, flow, and \Peclet number conditions, further developments are needed. Similarly to the reduced-order models widely adopted for cardiovascular flow problems\cite{Vignon-Clementel2006,Vignon-Clementel2010OutflowArteries,Arthurs2017ReproducingDesign}, it is critically important to develop reduced-order models of mass transport for the proximal and distal portions of the vascular system not included in the 3D geometric model. This is particularly important when dealing with closed-loop models and simulations involving reaction.

The DC scheme introduces a non-linear term in the weak form of the problem. Although an increase in computational cost was expected, no increase was observed in simulations run with the DC operator. This expense can be mitigated by the use of a time-lagging DC scheme\cite{Catabriga2002ImprovingEquations}. We remark, however, that in the presence of nonlinearity (e.g. in source terms), the \sad problem would be nonlinear regardless of the DC scheme and computational cost will not be significantly different with and without the DC operator.

\section{Acknowledgements}
SL is supported by the NSF Graduate Research Fellowship Program and the American Heart Association Fellowship (AHA 18PRE33960252). CAF is supported in part by the Edward B. Diethrich M.D. Professorship. NN is supported by the American Heart Association Fellowship (AHA 20POST35220004). CA is supported by the Wellcome Trust, and acknowledges The Centre for Medical Engineering (CME) at King's College London. ZX and OS gratefully acknowledge the financial support of NSF CAREER grant 1350454 and U.S. Army grant  W911NF-19-C-0094. Additionally, computing resources were provided by the NSF via grant no. 1531752 MRI: Acquisition of Conflux. 

\newpage
\bibliography{bibliography}
\end{document}